\documentstyle[12pt]{article}

\newcommand{\be}{\begin{equation}}
\newcommand{\ee}{\end{equation}}
\newcommand{\bea}{\begin{eqnarray}}
\newcommand{\eea}{\end{eqnarray}}

\begin{document}
\begin{titlepage}
\flushright{IP/BBSR/2000-18}
\flushright{hep-th/0005113 }

\vspace{1in}

\begin{center}
\Large
{\bf  Noncommutative Open String, D-brane and Duality  }

\vspace{1in}

\normalsize

\large{Jnanadeva  Maharana and Shesansu S.  Pal}
\it {e-mail:maharana@iopb.res.in and shesansu@iopb.res.in}

\normalsize
\vspace{.7in}

{\em Institute of Physics \\
Bhubaneswar - 751005, India }

\end{center}

\vspace{.5in}

\baselineskip=12pt
\begin{abstract}
\noindent 

 We consider open strings ending on D-branes in the presence of constant
metric, G,  antisymmetric tensor, B and gauge field, A. The Hamiltonian is
 manifestly invariant under a global noncompact group; strikingly similar to
toroidally compactified closed string Hamiltonian.
The evolution equations for the string coordinates, $\{ X^i \}$ and their
dual partners, $\{Y_i \}$, are combined to obtain equations of motion
invariant under the noncompact symmetry transformations. We show that
one can
start  from a noncommutative theory, with nonvanishing G and B and mixed
boundary conditions and then go over to a dual theory whose coordinates
obey Dirichlet boundary conditions. It is possible to generate B-field
by implementing the noncompact symmetry transformation. The connection
between this duality transformation and Seiberg-Witten map is discussed.

\end{abstract}

\end{titlepage}



The discovery of noncommutativity property in string theory and in field
theories has attracted considerable attention in recent times.
 When one compactifies the M(atrix) model on a
$T^2$ in the presence of antisymmetric tensor, noncommutative supersymmetric 
Yang-Mills theory appear naturally as
was shown by Connes, Douglas and Schwarz \cite{conds}. This can be 
understood from the D-string point of view if one dualizes one cycle of
the torus \cite{md}. Subsequently, the origin of the noncommutativity 
property from the perspective of D0-branes have been studied by
several authors \cite{ckr,ko}. Another interesting result is that the D-brane
worldvolume exhibits noncommutativity  in  string theory 
even before one goes
over to the M(atrix) model limit \cite{hv,naty}. 
Furthermore, there have been attempts to
explain noncommutativity on D-brane world volume through the study of 
open string quantization in the presence of background fields 
\cite{chco,rest1,rest2,rest3}.
Recently, Seiberg and Witten \cite{ne}have provided deeper insight into the
 relation between string theory and  noncommutative geometry  and they 
have investigated various aspects of nonabelian
gauge theories in this context.\\
 Recent progress in string theory has enriched our  understanding of the
dynamics of string theory and has revealed  interconnection between the
five perturbatively distinct theories \cite{rev1}. 
It is recognized that
dualities have played a cardinal role in these developments 
\cite{givr,sen1,maha}. 
 The discovery of Dp-branes has
opened up investigations in new directions in string theories \cite{jopol} 
and we can visualize
these objects as spacetime hypersurfaces on which open strings can end. 
Consequently, one can explore various aspects of string dynamics and
establish connections between string theory and supersymmetric
Yang-Mills theories from a novel perspective.It has been conjectured
that strongly coupled spatially noncommutative $ {\cal N} =4 $ SYM has a
dual description in terms of open string theory in a near critical
electric field\cite{gmms}. Some aspects of T-duality, in the context of
Morita equivalence, have been studied \cite{bs,ne} and the T-duality properties
of open strings in the presence of B-field have been considered from
the point of view of canonical transformations \cite{smj}. However, as we shall
see below, we utilize duality transformation which is analog of $O(d,d)$
transformation and such transformations  have interesting consequences. 
It is natural to expect that the brane world might reveal interesting
attributes when we examine the properties of its spacetime geometry. 
Where one   quantizes  an open string ending on a D-brane
with a constant antisymmetric field background (coming from NS-NS sector),
then it can be shown that spacetime coordinates of the open string
endpoints are noncommutative.\\ 
The purpose of this investigation  is to explore the 
consequences of target space
duality for the configurations when open strings end on D-branes and
there is constant background metric, $G_{ij}$ and NS-NS antisymmetric tensor
field, $B_{ij}$. We intend to  explore a scenario where one starts  from
a theory with noncommuting string coordinates and then go over to a dual
set of coordinates and backgrounds and examine whether the dual theory is
a noncommutative or commutative one. 
We shall also consider the situation when the two end points
of the open string are attached to the same brane and we couple string 
to the resulting $U(1)$ gauge field with constant field strength, i.e.
$F=dA$ is constant. We  show that the Hamiltonian obtained from the
worldsheet action can be cast in a form similar to the one derived
for closed strings for constant backgrounds exhibiting $O(d,d)$ invariance.
We argue that there is an analogue of the $O(d,d)$ symmetry in this situation.
It is well known that if one has an open string where some of the coordinates 
obey  the Dirichlet boundary conditions and the rest of them satisfy  
Neumann boundary conditions, then the usual T-duality operation interchanges 
these
boundary conditions to one another. We also study  how the boundary
conditions are modified under  the duality transformations
mentioned  above.\\
Let us consider, for the sake of definiteness,  a fundamental bosonic string 
ending on a Dp-brane, 
        although our considerations are applicable to 
 type II superstring or type 0 superstring.
The action \cite{a1,a2}, in the orthonormal gauge,  is given by  
\bea 
\label{action}
S=-{1\over 2} \int d^2 \sigma (\gamma ^{\alpha\beta} 
G_{ij}\partial _{\alpha}X^i \partial_{\beta}
X^j +\epsilon ^{\alpha \beta}\partial _{\alpha}X^i \partial _{\beta}X^j B_{ij})
- \int _{\partial \Sigma}
d\tau A_i \partial _{\tau} X^i\eea
where $\gamma ^{\alpha\beta} = {\rm diag}(-1~1)$, $\epsilon ^{01} =1,
~ \epsilon ^{10}=-1 $ and $\{X^i \},~i=0,1...p $
are the coordinates of the string along the brane directions. 
The constant nonvanishing backgrounds are: 
metric, $G_{ij} $ and  antisymmetric tensor, 
$B_{ij}$, $i,j =0,1..p$, with H=dB=0,
The rest of the coordinates are
denoted as ${X^a}, a=p+1,..D$ and we have not written the action for these
coordinates here. Moreover,
we envisage the configuration when the two ends of the open string are attached
to the same Dp-brane and $A_i$ is the resulting U(1) 
gauge field living on the world
volume of the brane. The last term in eq.(\ref{action}) is the coupling 
of the gauge 
connection  to the string. We could  rewrite this piece  
as ${1\over 2} \int _{
\Sigma} d^2\sigma \epsilon ^{\alpha \beta} F_{ij}\partial _{\alpha} X^i
\partial _{\beta} X^j $, where field strength, $F_{ij}$ is taken as  
constant.
As a consequence, (\ref{action}) can be expressed as
\bea
\label{actionII}
S=-{1\over 2}\int d^2\sigma[\gamma ^{\alpha\beta}G_{ij}\partial _{\alpha}X^i
\partial _{\beta} X^j +
\epsilon ^{\alpha\beta}\partial _{\alpha}X^i\partial _{\beta} X^j B_{ij} -
\epsilon ^{\alpha\beta}F_{ij}\partial _{\alpha}X^i\partial _{\beta} X^j] \eea
The resulting equation of motion is 
\bea
\partial _{\alpha }(\gamma ^{\alpha \beta}G_{ij}\partial _{\beta} X^j-
\epsilon ^{\alpha\beta} [F_{ij}-B_{ij}]\partial _{\beta}X^j )=0 \eea  
 the boundary conditions at $\sigma =0,\pi $ are 
\bea 
\label{bc}
G_{ij}\partial _{\sigma} X^j - {\cal F}_{ij}\partial _{\tau} X^j =0 
\eea  
where ${\cal F}_{ij} =B_{ij}-F_{ij}$. The noncommutativity character of the
string coordinates is revealed as follows. First   the 
worldsheet coordinates are complexified
through the definition $z=\tau +i\sigma$, after $\tau$ has been rotated 
to Euclidean signature. Next,  one maps the disc geometry of the
openstring to the upper half z-plane and then computes the correlation function
$<X^i(z)X^j(z')>$ with the above boundary condition (\ref{bc}), which has
the following form. \cite{ne}
\bea
\label{noncom}
\langle X^{i}(z)X^{j}(z^{'})\rangle =-\frac{1}{2\pi}[
G^{ij}\log|z-z^{'}|-G^{ij}\log|z-{\bar z}^{'}|+\bar{G}^{ij}\log|z-{\bar z}
^{'}|^{2}
+\theta^{ij}\log\frac{z-{\bar z}^{'}}{\bar{z}-z^{'}}]
\eea
 Where $ \bar{G}^{ij} = [(G+{\cal F})^{-1}G(G-{\cal F})^{-1}]^{ij} $ and
$ \theta^{ij} = -[(G+{\cal F})^{-1}{\cal F}(G-{\cal F})^{-1}]^{ij} $
The above result can be  derived in a straight forward manner following
the work of Abouelsaood et al \cite{acny}.
It is of interest to consider the propagator on the boundary which has the
disc geometry

\begin{equation}
\label{greenf}
<X^i(\tau)X^j(\tau ')>=-\frac{1}{2\pi}{\bar{G}}^{ij}\log (\tau -\tau ')^2+{i\over 2}\theta
^{ij}\epsilon (\tau -\tau ')
\end{equation}
where $\epsilon (\tau)$ takes values $+1$ and $-1$ for positive and negative
argument respectively. Note that ${\bar{G}}^{ij}$ is the effective metric
seen by the open string. Moreover, this also determines the anomalous
dimension of the vertex operator as this metric appears as the coefficient of
the $\log (\tau -\tau ')^2$. Note that the noncommutativity
property of the spatial coordinates can be brought out starting from
the correlation functions of the string coordinates and then defining
them on the real z-axis.
The noncommutativity feature of the coordinates ${X^i}$ is derived
from the structure of the correlation function.

However, at this stage, we proceed to derive the Hamiltonian associated with
the action (\ref{actionII}), keeping Lorentzian signature for $(\tau ,\sigma)$.
The canonical momenta are  given by
\be P_i=G_{ij}{\dot X}^j - {\cal F}_{ij}X'^j \ee
note that dot and prime refer to the derivatives with respect to $\tau$ and
$\sigma$ here and everywhere. 
 The Hamiltonian density is given by
\bea
\label{ham}
H={\frac{1}{2}}\left( \begin{array}{cc}  P_i &  X'^i \end{array} \right) 
{\cal M}_{ij} \left( \begin{array}{cc} P_j \\ X'^j \end{array} \right)
\eea
where 
\bea
{\cal M}_{ij}  = \left( 
\begin{array}{cc} G^{ij} & G^{ik}{ \cal  F}_{kj} \\
- {\cal  F}_{ik}G^{kj} & G_{ij}- {\cal  F}_{ik}G^{kl} 
{\cal F}_{lj} \end{array} 
\right)
\eea
Notice the  form of the Hamiltonian with $\cal M $-matrix. This has
striking similarity with the expression one derives for the
Hamiltonian that appeared for closed string
with constant background metric and antisymmetric field ($B_{ij}$) where
the corresponding M-matrix appears with specific combinations of the 
backgrounds \cite{grv}.
 In the present context, we note the appearance of $ {\cal F}$
in the the  ${\cal M}$-matrix which  has replaced the -B-field
of the M-matrix. 
We may argue that with the appearance of symmetric $\cal M$-matrix and the
combination of $P_i ~{\rm and}~ X'^i $ in the expression (\ref{ham}) implies 
that the Hamiltonian is $O(p+1,p+1)$ invariant; if we demand
the transformations 
\bea \left(\begin{array}{cc} P_i \\ X'^i \end{array}\right) \rightarrow (\Omega
^{-1})_{ij} \left(\begin{array}{cc} P_j \\ X'^j \end{array}\right) ~~~ {\cal M}
\rightarrow \Omega ^T{\cal M} {\Omega }~~~ {\rm and}~~~
 \Omega \eta \Omega ^T =
\eta \eea   
Here $\eta =\left(\begin{array}{cc} 0 & {\bf 1} \\ {\bf 1} & 0 \end{array}
\right)$ is the metric of the $O(p+1,p+1)$ group, $\bf 1$ being $(p+1)\times 
(p+1)$ unit matrix. $\Omega $ is,  $(2p+2) \times (2p+2)$
matrix,  which is an arbitrary element of the  the global $O(p+1,p+1)$ group.
Let us set the gauge potential $A_i =0$ from now on, for the sake of 
simplicity.\\
  A few comments are in order at this stage. In the case of toroidal 
compactification, when d-spatial coordinates are compactified
on $T^d$    
\cite{nsw} the
T-duality group is $O(d,d,Z)$ and this is a generalization of the $R\rightarrow {1\over
R}$ duality appearing in the case of compactification on $S^1$.  As is well 
known, if we consider a theory with a given set of background fields which
is described by a conformal field theory, these duality transformations can
take us to another set of background fields and we can have a conformal
field theory for the new set.\\
When we consider open string theories with  some of the coordinates
obeying  Neumann boundary condition and the rest of the set fulfilling Dirichlet
boundary conditions, the simplest form of T-duality (i.e. analogue
$R\rightarrow {1\over R}$ transformation) interchanges the two types of 
boundary conditions (actually it is $P\leftrightarrow  X'$ transformation 
for trivial 
backgrounds). Therefore, in view of the appearance of afore mentioned
$O(p+1,p+1)$ symmetry, we expect that starting from a constant $G_{ij}$
we can generate both G and B with a judicious choice of $\Omega $ matrix.
Thus, one might consider, to begin with,   background configurations 
with constant $G_{ij} ~{\rm and }~ B_{ij}$ and choose the boundary condition
$G_{ij}X'^j - B_{ij}{\dot X}^j =0,   
~ {\rm at} ~ \sigma =0, \pi $. In this case,correlation functions of  the string coordinates $X^i(z)
~{\rm and }~ X^j(z') $ are given by $(5)$. Then by going over to a
dual theory, we shall find that the coordinates of that theory
are commutative. \\ 
In order to proceed in  this direction we need to consider the 
underlying T-duality
symmetry while studying evolution equations of the worldsheet coordinates.
 This is revealed in
an elegant manner \cite{mike,jm,ms} by introducing auxiliary fields,
 $U^i_{\alpha}$,  
described below. Let us consider the Lagrangian density
\bea
\label{l1}
{\cal L}_1=
-{1\over 2}\gamma ^{\alpha\beta}U^i_{\alpha}U^j_{\beta} G_{ij} -{1\over 2}
\epsilon ^{\alpha\beta}U^i_{\alpha}U^j_{\beta}B_{ij} +\partial _{\alpha} X^i(
\gamma ^{\alpha\beta}U^j_{\beta}G_{ij}+\epsilon ^{\alpha\beta}U^j_{\beta}B_{ij})
\eea
The equations of motion associated with $U^j_{\beta}$ and $X^i$, 
respectively,  
 are
\bea
\label{equ1}
(\partial _{\alpha}X^i -U^i_{\alpha})(\gamma ^{\alpha\beta}G_{ij}+\epsilon ^{
\alpha\beta}B_{ij})=0 \eea
\bea
\label{equ2}
\partial _{\alpha}(\gamma ^{\alpha\beta} U^j_{\beta}G_{ij}+\epsilon ^{
\alpha\beta}U^j_{\beta}B_{ij})=0 \eea
If we solve for $U^i_{\alpha}$ from (\ref{equ1}) and substitute in (\ref{equ2})
we recover the equation of motion for $\{X^i \}$ coordinates at the classical
level. Now let us 
introduce a set of `dual' coordinates $\{Y_i \}$ and correspondingly auxiliary
fields $\{V^i_{\alpha} \}$ and another Lagrangian density 
\bea
\label{l2}
{\cal L}_2=
{1\over 2} \gamma ^{\alpha\beta}V^i_{\alpha}V^j_{\beta} G_{ij} 
+{1\over 2}\epsilon ^{\alpha\beta
}V^i_{\alpha}V^j_{\beta}B_{ij} + \epsilon ^{\alpha\beta}\partial _{
\alpha}Y_iV^i_{\beta} \eea
The equations of motion associated with $V^i_{\alpha}$ and $Y_i$ are
\bea
\label{eqv1}
\gamma ^{\alpha\beta}V^j_{\beta}G_{ij}+\epsilon ^{\alpha\beta}V^j_{\beta}B_{ij}
+ \epsilon ^{\beta \alpha} \partial _{\beta} Y_i =0 \eea
\bea
\label{eqv2} 
\partial _{\alpha}(\epsilon ^{\alpha \beta} V^i_{\beta})=0 \eea
We can express $V^i_{\alpha}$ in terms of $Y_i$ coordinates through the 
relation
\bea
V^i_{\alpha} = \gamma _{\alpha \beta} \epsilon ^{\beta \delta}P^{ij}\partial _{
\delta}Y_j+Q^{ij}\partial _{\alpha}Y_j \eea
where 
\bea
\label{GBPQ}
P=B^{-1}(GB^{-1}-BG^{-1})^{-1} ~~~{\rm  and} ~~~ 
Q=-G^{-1}(GB^{-1}-BG^{-1})^{-1}  \eea
are symmetric and antisymmetric tensors respectively.
\\
The equations of motion derived from (\ref{l1}) suggest that we can write
$U^{\alpha}_i=\epsilon ^{\alpha\beta}\partial _{\beta}Y_i $, at least locally,
 and
similarly we may conclude, from  equations of motion, associated with 
${\cal L}_2$, for $Y_i$ coordinates that  
\bea
\label{Y1}
{{\partial {\cal L}_2}\over {\partial ({\partial _{\alpha}} Y_i)}}= 
\epsilon ^{\alpha\beta}\partial _{\beta} X^i \eea
 Therefore, we have two sets of local relations 
\bea
\label{Y}
\epsilon ^{\alpha\beta} \partial _{\beta}Y_i =\gamma ^{\alpha\beta}
\partial _{\beta}
X^jG_{ij} + \epsilon ^{\alpha\beta}\partial _{\beta} X^jB_{ij} \eea
\bea 
\label{X}
\epsilon ^{\alpha\beta}\partial _{\beta} X^i= \gamma ^{\alpha\beta} 
\partial _{\beta}Y_iP^{ij}+\epsilon ^{\alpha\beta}
\partial _{\beta }Y_jQ^{ij} \eea
If we examine the field equations for $\{X^i \}$ coordinates, it is the
divergence of the $r.h.s$ of (\ref{Y}). But this is nothing but the Bianchi
identity for the dual coordinates $Y_i$. The same
is true when we consider (\ref{X}). This is all 
very familiar whenever we consider
such dualities; the field equations for one set is the Bianchi identity for
the dual variables and vice versa. 
An important point deserves to be mentioned {\it en passant}:  these 
two evolution 
equations can be expressed in a simple and  elegant form \cite{mike,jm,ms} 
if we enlarge the manifold where $\{X^i \}$ and $\{ Y_i \}$ are treated as
independent coordinates. Let W stands for the $2(p+1)$ coordinates $\{X^i,Y_i \}
$ collectively, then equations (\ref{Y}) and (\ref{X}) can be expressed as
$ {\cal M}\eta\partial _{\alpha}W=\epsilon _{\alpha\beta}\gamma ^{\beta\delta}
\partial _
{\delta} W $. 
The evolution
equation becomes
$\partial _{\alpha}(\eta {\cal M}^{-1}\gamma ^{\alpha\beta}
\partial _{\beta} W)=0 $
which is $O(p+1,p+1)$ invariant.
\\
Now let us focus attention on  ${\cal L}_2 $, which can be expressed in 
the following form after  eliminating 
auxiliary fields $V^i_{\alpha}$ in favor of coordinates   $Y_i$ 
through (17).
\bea
\label{l2Y}
{\cal L}_2= {1\over 2}[ \gamma ^{\alpha\beta}\partial _{\alpha}Y_i\partial _{
\beta}Y_jP^{ij} +\epsilon ^{\alpha\beta} \partial _{\alpha} Y_i
\partial _{\beta}Y_jQ^{ij}] \eea
The resulting equation motion, in the presence of constant 
backgrounds $P^{ij}~{\rm and}
~ Q^{ij}$,  are
\bea
\label{eqmY}
\partial _{\alpha}(\gamma ^{\alpha\beta}\partial _{\beta} Y_jP^{ij} +
\epsilon^{\alpha \beta}\partial _{\beta}Y_jQ^{ij})=0 \eea
with the   boundary condition
\bea
\label{bcY}
P^{ij}Y'_j-Q^{ij}{\dot Y}_j =0 \eea
We recall that, in the presence of the antisymmetric background $Q^{ij}$,
the coordinates $\{Y_i \}$ will exhibit noncommutativity property. In order
to demonstrate this attribute, one needs to go over to Euclidean worldsheet
description and introduce complex coordinates $(z, {\bar z})$ as mentioned 
earlier.\\ 
Let us  now, utilize the relations between $({\dot Y}_i,
Y'_i) ~{\rm and} ~ ({\dot X}^i, X'^i)$, from (\ref{Y}) and (\ref{X}), to write 
the mixed boundary condition (\ref{bcY}) in terms of the $\{X^i \}$ 
coordinates. Thus we   arrive at
\bea
P^{ij}(-G_{jk}{\dot X}^k+B_{jk} X'^k)-Q^{ij}(-G_{jk} X'^k+B_{jk}{\dot X}^k)=0
\eea
Now substitute the relations between $P^{ij}, Q^{ij}, G_{ij} ~{\rm and}~
B_{ij}$ using (\ref{GBPQ}) in the above equation which leads to
\bea  
\label{Nbc}
{\dot X}^j=0 \eea
We have shown that if one starts from a  theory with constant metric and 
antisymmetric
tensor field with mixed boundary conditions (\ref{bcY}), so that one
has a noncommutative theory;  it is
possible to go over to a theory described by a set of dual coordinates,
with a (corresponding) constant metric, satisfying Dirichlet 
 boundary conditions
(\ref{Nbc}) and   consequently, the dual theory is a commutative one.\\
Next, we address another question in this context. Suppose, we start with
a commutative theory such that $G_{ij}$ are constant and $B_{ij}$ are 
set to zero
with Neumann boundary condition i.e. $G_{ij}X'^j=0$; then,  is it possible to
generate a B-field and a mixed boundary condition? Our  answer is affirmative
in this regard as is  illustrated  by the following  example.
\\
Lets us  consider infinitesimal $O(p+1,p+1)$ transformation introduced by
Maharana and Schwarz \cite{ms} where
$\Omega $ has the following form
\bea
\label{inf}
\Omega=\left(\begin{array}{cc} {\bf 1}+\alpha & \beta \\
\lambda & {\bf 1} - \alpha ^T \end{array}\right)
\eea
Here,  $\alpha, \beta ~{\rm and}~ \lambda$ are infinitesimal parameters (actually
$(p+1)\times(p+1)$ matrices). The constraint $\beta ^T=-\beta$ and $\lambda ^T=
-\lambda  $ follow from the condition  
$\Omega \eta \Omega ^T =\eta$.   
Let us consider an initial configuration 
\bea 
{\cal M} = \left(\begin{array}{cc} {G^{-1}} & 0\\
0 & G \end{array} \right) \eea
and implement the transformation (\ref{inf}), then we find that
\bea 
G_{ij}  \rightarrow {\tilde G}_{ij} = G_{ij} -G_{il} (\alpha ^T)^l_j -
{\alpha  }^l_iG_{lj} \eea 
We have retained only linear terms in the infinitesimal matrices in the 
above equation. Note that the pair of 'phase space variables' $\{P_i, X'^i \}$
will also get transformed according to (9) and ${\tilde X}'^i =
-\lambda  ^{ij}P_j+{(1+\alpha ^T  )}^i_j X'^j $. Therefore, 
keeping only the linear
terms in infinitesimal parameters, we arrive at the following form of 
mixed  boundary conditions,
\bea
\label{mixed}
-G_{ij}\lambda ^{jk}G_{kl}{\dot X}^l +(G_{ij}-\alpha ^k_iG_{kj})X'^j =0 \eea
using the definition $P_i=G_{ij}{\dot X}^j $. In (\ref{mixed}) we have made
a special choice for the infinitesimal parameter $\alpha ^{ij}$ is symmetric
and is proportional to $\delta ^{ij}$.   

Notice that we have generated a constant antisymmetric tensor field $B_{ij}=
G_{ik}\lambda ^{kl}G_{lj}$, through a T-duality transformation
by choosing a specific form of the  matrix
$\Omega$, $\lambda$ being the infinitesimal matrix valued parameter. As is
well known, there are distinctions between a theory where $B=0$ and the one
with constant, nonzero B-field for the situation under consideration. Let
us introduce the physical vertex operator for gauge field: $V_A=\int \xi .
\partial Xe^{ip.X}$, where $\xi _i $ is the polarization vector and for
the physical vertex operator the constraints are $\xi .p=0 ~{\rm and}~ p^2=0$.
Recall that all contraction here are done with open string metric ${\bar{G}}_{
ij}$. For the  $B=0$ case, when the correlations of vertex operators are
computed using (\ref{greenf}), there is no $\theta$ dependence;
in other words we are dealing with ordinary gauge fields. On the
other hand, when B-field is nonzero, there will be $\theta$ dependent
terms if we calculate correlations of gauge field vertex operators. In fact
the product of the vertex operators in this case is the $*$ product with
the identification of $\theta ^{ij}$ as the noncommutative parameter of
the $*$ product. Let us recall briefly, the gauge field dynamics from the
perspective of $\sigma$-model. The gauge field background term
$- \int _{\partial \Sigma}
d\tau A_i \partial _{\tau} X^i$ appearing in (\ref{action}) is invariant
under $\delta _{\psi}A_i=\partial _i\psi$, and $\psi$  being the
 gauge parameter. This
is valid at the classical level since the gauge variation goes to a total
derivative ensuring gauge invariance. However, in the field theoretic frame
work, a regularization prescription needs to be adopted while computing
the variation. If the Pauli-Villars regularization is chosen the corresponding
theory is an ordinary gauge theory whereas for point splitting regularization
the gauge transformation required is that of noncommutative theory as has
been argued by Seiberg and Witten \cite{ne}. We also mention that in case
of the Pauli-Villars regularization, when one constructs effective action,
its B and F dependence appears in the combination of $B+F$ and the gauge
invariance is guaranteed due to the transformation properties: $\delta B=
d\Lambda ~{\rm and} ~ \delta A=\Lambda$, $\Lambda$ being a one form. Therefore,
the effective action one gets is intimately connected with the choice of
regularization prescription in the context of the $\sigma$-model: Pauli-Villars
regularization leads to ordinary gauge theory and point splitting to the
noncommutative one. Let us adopt the notation $A$ and $\hat A$ for the
ordinary and noncommutative gauge fields respectively
and similarly for the gauge parameters following \cite{ne}.
The Seiberg-Witten map provides means to express $\hat A$ in terms of $A$
and $\hat {\psi}$ in terms of $A$ and $\psi$. If one starts with a theory
with field strength $\hat F$ and noncommutative parameter $\theta$, then
the relation is
\begin{equation}
\label{swmap}
{\hat F}={1\over {1+F\theta}}F
\end{equation}
where $F={\hat F}(\theta =0)$. Therefore, an ordinary gauge field with constant
curvature and NS two form field B is equivalent to noncommutative gauge
field theory with the appropriate$\theta$-parameter.
 Thus it follows from (\ref{swmap})
that,for $\theta=\frac{1}{B}$
\begin{equation}
\label{swmapb}
{\hat F}=B{1\over {{B+F}}}F
\end{equation}
It is obvious from (\ref{swmap}) and (\ref{swmapb}) that $\hat F$ will blow
up at some points and same argument will go through when we express F in
terms of $\hat F$ and $\theta$ by inverting the equation.\\
In our work, we are able to generate a constant B-field through duality
transformations. Therefore, we can construct the Seiberg-Witten map explicitly
in terms of the parameter of the noncompact symmetry transformation. The
two field strengths are now related by the expression
\begin{equation}
\label{oddmap}
{\hat F}=G\lambda(G\lambda +FG^{-1})^{-1}F
\end{equation}
and the matrix multiplications are obvious in the above formula.
\\
We know that noncommutative field theories have the unusual feature that
UV and IR cutoffs get interlinked as has been discussed recently \cite{rvn}
. The point
is IR diverges appear from the contribution of the nonplanar diagrams when
the momenta of external particles go to zero. It is  argued that
the star product appearing in the action of noncommutative theories are
neither maximally Lorentz invariant nor local. The origin of nonlocality
can be traced to the fact that in defining the star products one introduces
an infinite number of derivatives (note however, that the quadratic part
of the action remains same for ordinary product of fields or star products).
If we look at two noncommuting spatial coordinates i.e.
$[X^i,X^j]=1\theta ^{ij}$, it can be argued that a short distance scale in
one coordinate, say $X^i$ corresponds to a long distance regime in $X^j$, since
noncommutativity gives rise to a built in uncertainty relation $\Delta X^i
\Delta X^j \ge {1\over 2} |\theta ^{ij} | $.
Now, for our case same issue
can be raised. From string theory point of view, if one considers,
scattering of gauge bosons then there will be oneloop open string
diagram which one would have to consider. But the same diagram from the 
closed string point of view is a tree diagram. In fact the
relation between noncommutative and ordinary gauge theory processes can be
viewed in a more transparent way, if one considers the zero slope limit
and then scales metric, $\alpha '$ and the B-field appropriately as is
explained in \cite{ne}. However, we are not in a position to go over to
zero slope limit and adopt the scaling of \cite{ne} in a straight
forward manner, since our B-field is generated by the infinitesimal
transformation. Nevertheless,  the nonlocality picture should be viewed from
the point of view of closed and open string diagrams as mentioned above.

To summarize, one of our interesting results is that we started with a theory
endowed with noncommutative string coordinates since the
constant backgrounds G and B are nonvanishing and  
  the mixed boundary conditions. Then we 
showed that there exist a set of dual coordinates and backgrounds such that
 these coordinates satisfy Dirichlet boundary conditions and these dual 
coordinates belong to a commutative theory. Thus there is an interesting 
interconnection between a theory which is endowed with the
noncommutative and a commutative theories.

 We presented another example to show that the infinitesimal
 noncompact symmetry transformation $\Omega$ generates mixed boundary 
condition for a special choice of the parameter. We also presented
arguments to establish relation between the parameters of the duality
transformation and Seiberg-Witten map. 
\begin{center}
{\bf Acknowledgement}\\
\end{center}
We are thankful to the anonymous
referee for several constructive suggestions.
We would like to thank the participants of the extended workshop held at
MRI in 1999.



\vspace{.7in}
\begin{center}
{\bf References}
\end{center} 
\begin{enumerate}


\bibitem{conds} A. Connes, M. R. Douglas and A. Schwarz, JHEP {\bf 9802}
(1998) 003,hep-th/971112.
\bibitem{md} M. R. Douglas and C. Hull, JHEP {\bf 9802}
(1998) 008, hep-th/971165.
\bibitem{ckr} Y. -K. E. Cheung and M. Krog, Nucl. Phys. {\bf B 528} (1998) 185,
hep-th/9803031.
\bibitem{ko} T. Kawano and K. Okyuyama, Phys. Lett. {\bf B433} (1998) 29,
hep-th/9803044.
\bibitem{hv} C. Hofman and E. Verlinde, JHEP 9812 (1998) 010. hep-th/9810116.
\bibitem{naty} N. Seiberg, Talk given at the New Ideas in Particle Physics
and Cosmology, Univ. of Pennsylvania, May 19-22, 1999.
\bibitem{chco} C.-S. Chu and P.-M. Ho, Nucl. Phys. {\bf B550} (1999),
hep-th/9812219; Nucl. Phys. {\bf B568} (2000) 503.
\bibitem{rest1} V. Schomerus, JHEP {\bf 9906} (1999) 030.hep-th/9903205
\bibitem{rest2} F. Ardalan H. Arfaei and M. M. Sheikh-Jabbari, 
Dirac Quantization of Open Strings and Noncommutativity in Branes, 
hep-th/9906161.
\bibitem{rest3} N. Ishibashi, A Relation between Commutative and Noncommutative
Description of D-branes, hep-th/9909176.
\bibitem{ne} N. Seiberg and E. Witten, String Theory and Noncommutative
Geometry, hep-th/9908142.
\bibitem{rev1}  J. H. Schwarz, Superstrings and M TheoryDualities, 
TASI Summer School Lectures,
hep-th/9607201.
\bibitem{givr} A. Giveon, M. Porrati and E. Rabinovici, Phys. Rep. {\bf C244}
(1994) 77-202.hep-th/9401139
\bibitem{sen1} A. Sen, An Introduction to Non-perturbative String Theory,
hep-th/9802051.
\bibitem{maha} J. Maharana, Recent Developments in String Theory,
hep-th/9911200.
\bibitem{jopol} J. Polchinski, TASI Lectures on D-branes,
  hep-th/9611050.
\bibitem{gmms} R.Gopakumar, S.Minwalla, J.Maldacena and
  A.Strominger, hep-th/0005048. and  N.Seiberg, L.Susskind and
  N.Toumbas, hep-th/0005040 for similar considerations.
\bibitem{bs} B. Pioline and A. Schwarz, Morita Equivalence and T-duality
(or B versus $\theta $), hep-th/9908019.
\bibitem{smj} M. M. Sheikh-Jabbari, A note on T-duality, Open strings in
B-field Backgrounds and Canonical Transformations, hep-th/9911203.
\bibitem{a1} J. Dai, R. G. Leigh and J. Polchinski, Mod. Phys. Lett. {\bf A4}
(1989) 2073.
\bibitem{a2} R. Leigh, Mod. Phys. Lett. {\bf A4} (1989) 2767.
\bibitem{acny} A. Abouelsaood, C. G. Callan, C. R. Nappi and S. A. Yost,
Nucl. Phys. {\bf B280} (1987) 599.
\bibitem{grv} A. Giveon, E. Rabinovici and G. Veneziano, Nucl. Phys. {\bf B322}
(1989) 167; A. Shapere and F. Wilczek, Nucl. Phys. {\bf B320} (1989) 669.
\bibitem{nsw} K. S. Narain, H. Sarmadi and E. Witten, Nucl. Phys. {\bf B279}
(1987) 369.
\bibitem{mike} M. J. Duff, Nucl. Phys. {\bf B335} (1990) 610.
\bibitem{jm} J. Maharana, Phys. Lett. {\bf B 296} (1992) 65-70.hep-th/9205016
\bibitem{ms} J. Maharana and J. H. Schwarz, Nucl. Phys. {\bf B390} (1993) 3-32.hep-th/9207016
\bibitem{rvn} S.Minwalla,M.V.Raamsdonk and
 N.Seiberg,hep-th/9912072;M.V.Raamsdonk and N.Seiberg,hep-th/0002186

\end{enumerate}
\end{document}